\documentclass[10pt]{article}
\usepackage[top=0.85in,left=2.75in,footskip=0.75in]{geometry}

\usepackage{amsmath,amssymb}

\usepackage{changepage}

\usepackage[utf8x]{inputenc}
\usepackage[document]{ragged2e}

\usepackage{textcomp,marvosym}

\usepackage{cite}

\usepackage{nameref,hyperref}

\usepackage[right]{lineno}

\usepackage{microtype}
\DisableLigatures[f]{encoding = *, family = * }

\usepackage[table]{xcolor}

\usepackage{array}

\usepackage{pdflscape}
\usepackage{multirow}
\usepackage{longtable}
\usepackage{verbatim}
\newcolumntype{+}{!{\vrule width 2pt}}

\newlength\savedwidth

\raggedright
\usepackage[aboveskip=1pt,labelfont=bf,labelsep=period,justification=raggedright,singlelinecheck=off]{caption}

\makeatletter
\renewcommand{\@biblabel}[1]{\quad#1.}
\makeatother

\usepackage{lastpage,fancyhdr,graphicx}
\usepackage{epstopdf}
\fancyheadoffset[L]{2.25in}
\fancyfootoffset[L]{2.25in}
\begin{document}
\vspace*{0.2in}

\begin{flushleft}
{\Large
\textbf\newline{Polygenic Risk Score in Africa Population: Progress and challenges} 
}
\newline
Yagoub Adam\textsuperscript{1\Yinyang},
Suraju Sadeeq\textsuperscript{2,3\Yinyang},
Judit Kumuthini\textsuperscript{6},
Olabode Ajayi\textsuperscript{6}, 
Gordon Wells \textsuperscript{6}, 
Rotimi Solomon\textsuperscript{1,3,4},
Olubanke Ogunlana\textsuperscript{1,3,4},
Emmmanuel Adetiba\textsuperscript{3,5,7},
Emeka Iweala\textsuperscript{3,4},
Benedikt Brors\textsuperscript{8,9},
Ezekiel Adebiyi\textsuperscript{1,2,3,9*}
\\
\bigskip
\textbf{1} Covenant University Bioinformatics Research (CUBRe), Covenant University, Ota, Ogun State, Nigeria
\\
\textbf{2}  Dept Computer \& Information Sciences, Covenant University, Ota, Ogun State, Nigeria
\\
\textbf{3}  Covenant Applied Informatics and Communication Africa Centre of Excellence (CApIC-ACE),  Covenant University, Ota, Ogun State,Nigeria
\\
\textbf{4}  Dept of Biochemistry, Covenant University, Ota, Ogun State, Nigeria
\\
\textbf{5}  
Dept of Electrical \& Information Engineering (EIE), Covenant University, Ota, Ogun State, Nigeria
\\
\textbf{6} Centre for Proteomic and Genomic Research, Cape Town, Western Cape, South Africa.
\\
\textbf{7} HRA, Institute for Systems Science, Durban University of Technology, Durban, South Africa.
\\
\textbf{8} German Cancer Consortium (DKTK), Heidelberg, Germany
\\
\textbf{9} Applied Bioinformatics Division, German Cancer Research Center (DKFZ), Heidelberg, 69120, Germany
\\
\bigskip

%
%
\Yinyang Joint first author.





* ezekiel.adebiyi@covenantuniversity.edu.ng

\end{flushleft}
\justify
\section*{Abstract}
Polygenic risk score (PRS) analysis is a powerful method been used to estimate an individual's genetic risk towards targeted traits.  PRS analysis could be used to obtain evidence of a genetic effect beyond Genome-Wide Association Studies (GWAS) results i.e. when there are no significant markers.  PRS analysis has been widely applied to investigate the genetic basis of several traits including rare diseases.  However, the accuracy of PRS analysis depends on the genomic data of the underlying population.  For instance, several studies showed that obtaining higher prediction power of PRS analysis  is challenging for non-Europeans. In this manuscript, we reviewed the conventional PRS methods and their application to sub-saharan Africa communities.  We concluded that the limiting factor of applying PRS analysis to sub-saharan populations is the lack of sufficient GWAS data.  Also, we  recommended developing African-specific PRS tools.
\\
\textbf{keywords}\\ Prediction medicine, GWAS, post-GWAS, PRS analysis, Africa population

\section*{Author summary}
PRS analysis is currently not applicable to African communities due to the current data sparsity with respect to Genome-Wide Association Studies (GWAS) in non-European ancestries. Therefore, extending the current PRS tools to handle diverse multi-ethnic data is crucial to estimate  PRS  values for personalized medicine across ethnic groups.



\section{Introduction}
Genome-wide association studies (GWAS) have been applied successfully to identify the associations between hundreds of genomic variations with complex human traits\cite{Bush2019}. In general, GWAS  report single nucleotides polymorphisms (SNPs) as statistically significant genomic variations associated with the trait of interest, when their \emph{p}-values are less than 5e-09 in Africa population (which also statistically depends on the number of SNPs analyzed)\cite{Gurdasani2019}. The statistically significant SNPs reported by GWAS are used to understand the biomolecular mechanisms of many phenotypic traits, including various human diseases. However, due to the statistical threshold, GWAS might fail to detect  SNPs that are associated  with low or moderate risks \cite{Cantor2010,Zhang2014}. The limitation of filtering  variants associated with low disease risk increases the GWAS false negative rate. Also, traditional GWAS cannot be used to integrate the  polygenic nature of  many complex traits \cite{Krapohl2017}.  Therefore, several post-GWAS approaches have  been introduced to overcome them \cite{Pasaniuc2016,Chimusa2019}.  Due to the privacy issues such as access to individual level GWAS data sets, most post-GWAS approaches require only GWAS summary statistics. There are many public resources for  GWAS summary  statistics and they include  the GWAS Catalog\cite{Buniello2019}, GWAS Central\cite{Beck2019} and the dbGaP database \cite{Mailman2007,Tryka2013}. A distinct  approach of performing post-GWAS analysis is known as PRS analysis. The PRS methods  map   genotype data from a GWAS summary  into a single variable used to estimate  an individual-level risk score for the phenotypic trait.

PRS analysis is used to predict an individual heritability by incorporating  all SNPs \cite{Dudbridge2013}. Therefore, obtaining a precise  PRS value from case-control studies can be used in personalized medicine.  However, challenges exist when translating PRS values to clinical care \cite{Lewis2017}. To successfully perform PRS analysis, two distinct GWAS summaries are required. The first sample (training sample) is used to select the SNPs for PRS analysis. The second sample (discovery sample) is used to evaluate the prediction value of PRS methods. The following PRS approaches  are discussed in the literature; (i) weighted methods that consider the  effect sizes derived from GWAS result, (ii) unweighted methods that consider the single marker analysis and (iii)  shrinkage methods  that consider the  multivariate analysis. In this review, we focused on the tools and methods that perform PRS analysis and  their applications in understanding the predictive power of PRS analysis. The reviewed PRS tools are chosen based on the following criteria:
\begin{enumerate}
    \item The approach must perform PRS analysis based on "base" (GWAS) data (summary statistics) and "target" data (genotypes and phenotypes in each of the target sample).
\item The approach may  involve linkage disequilibrium pruning.
\item The method or approach should be readily available in form of a tool or package to be able to execute the method on any data set.
\end{enumerate}

Besides reviewing the PRS methods,   we aim to investigate the application of PRS analysis on African population. It is important to note that the term "African population" includes all those whose ancestors are  African (i.e. Africans in diaspora). However, in this manuscript,   we will review the PRS studies only on sub-saharan Africa.
\section{Classification of PRS methods}
\begin{figure}[!h]
    \centering
    \includegraphics[scale=.25]{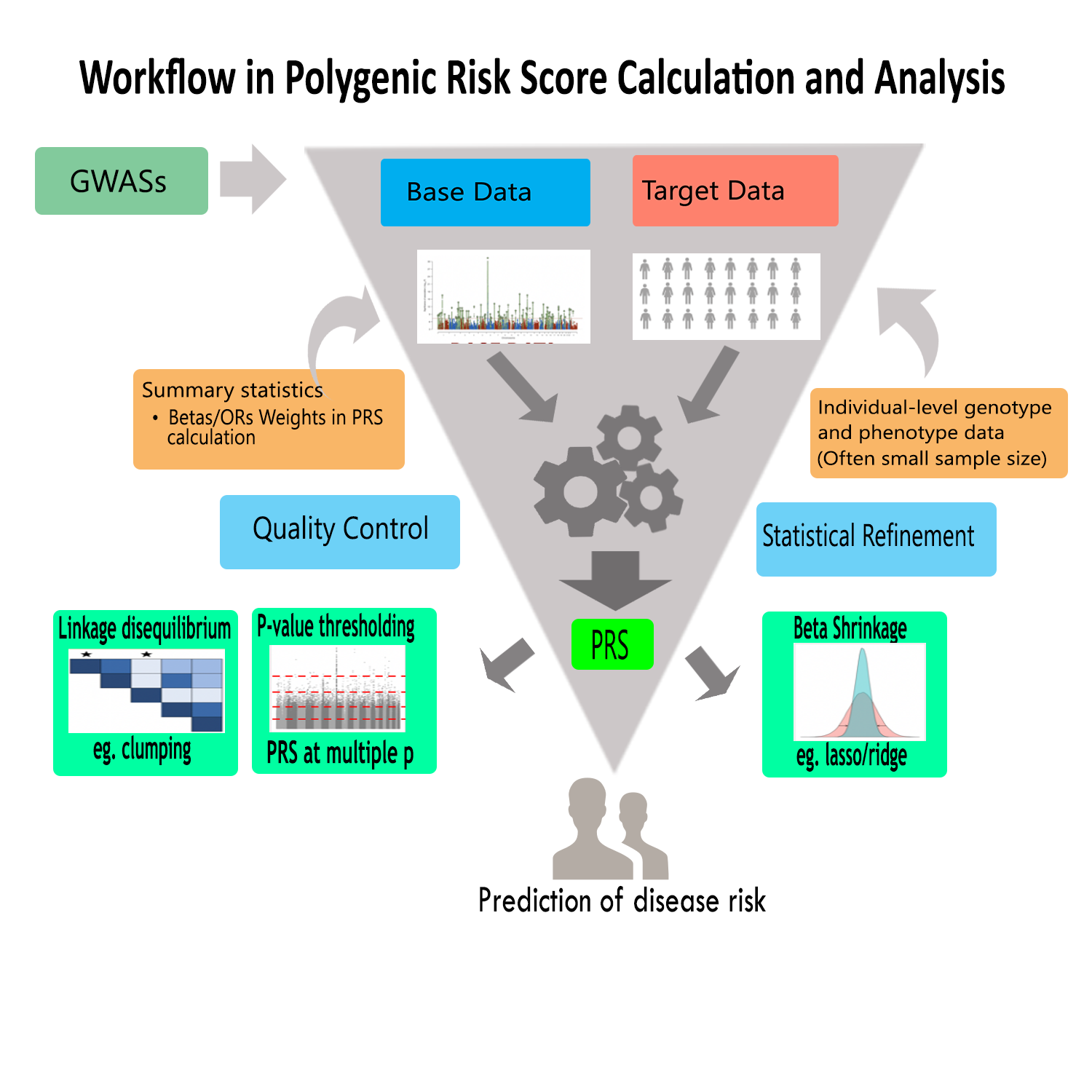}
    \caption{A general PRS analysis workflow. This is a typical polygenic risk score analysis workflow showing base data, target data and encapsulating different approaches. Using summary statistics and individual-level genotype and phenotype data, approaches such as lasso/ridge regression, clumping and \emph{p}-value thresholding can be employed to increase the predictive accuracy of PRS analysis. In addition, results may predict health or disease risk and provide insights for effective therapeutic interventions. }
    \label{Fig1}
\end{figure}

The different approaches under the umbrella of PRS analysis are presented in Figure \ref{Fig1} and Table \ref{table1}. We can categorize  PRS methods into two groups; (i) Bayesian-based methods (ii) non-Bayesian methods.  Also, we can classify  PRS methods into two with respect to their treatments of linkage disequilibrium:  (i) PRS methods that incorporate linkage disequilibrium (LD), and  (ii) PRS methods that apply LD pruning. However, to ease the understanding of their underlying algorithms, we grouped the PRS analysis approaches into four (See Table \ref{Tab2}):
\begin{enumerate}
    \item Clumping with thresholding (C+T) is the standard approach of polygenic scores analysis.
    \item \emph{p}-value thresholding approach.
    \item Penalized regression approach.
    \item Bayesian shrinkage  approach.
\end{enumerate}
\subsection{PRS methods that incorporate LD}
In practice, the prediction accuracy of PRS analysis tends to improve if the markers are LD pruned. However, as noted in Chatterjee  \textit{et al} \cite{Chatterjee2013}, the absence of linkage disequilibrium (LD) limits the predictive accuracy of PRS analysis. In addition, a simulation test performed by Vilhjálmsson    \textit{et al} \cite{Yang2015} shows that in the presence of LD, the prediction accuracy of the PRS analysis widely used approach of LD pruning followed by \emph{p}-value thresholding   (P + T) under-predicts the heritability explained by the SNPs. One special method that incorporated LD in its study is  LDPred (Subsection \ref{LDPred}), a Bayesian approach in the presence of LD. If loci are to be linked, then the posterior mean effect can be derived analytically under a Gaussian infinitesimal prior. An arguably more reasonable prior for the effect sizes is a non-infinitesimal model where only a fraction of the markers are causal. For this reason, consider the following Gaussian mixture prior:

\begin{eqnarray}
     \beta\sim iid \begin{cases} 
     N\left(0, \frac{h^{n}_{g}}{M_{p}} \right) & \text{ with probability p} \\
    0  & \text{ with probability (1-p)}
   \end{cases},
\end{eqnarray}
where $p$ is the probability that a marker is drawn from a Gaussian distribution i.e. the fraction of causal marker. Similarly, from this model the posterior mean can be derived as 
\begin{eqnarray}
E\left(\frac{\beta_{i}}{\widetilde{\beta}^{l}}, D\right) \approx \left(\frac{M}{N h^{2}_{g}} I + D_{i} \right)^{-1}\widetilde{\beta}^{l},
\end{eqnarray}
where $D_i$ denotes the regional LD matrix within the region of LD and $\widetilde{\beta}^{l}$ denotes the least-squares estimated effects within that region. The approximation assumes that the heritability explained by the region is small and that LD with SNPs outside of the region is negligible
\subsection{PRS methods that  apply LD pruning}
These PRS methods are  non-Bayesian approaches that apply informed LD pruning (LD clumping) (figure \ref{Fig1}). Moreover, these methods are referred to as pruning and thresholding (P+T) PRS methods. For instance, using a univariate regression coefficient  ($r^2$)  with a threshold of 0.2, we could apply \emph{p}-value thresholding. The \emph{p}-value thresholding process is optimized over a grid concerning prediction accuracy in the validation data. LD pruning that preferentially prunes the less significant marker could yield more accurate predictions than pruning random markers. For the \emph{p}-value selection threshold,  researchers should include only  SNPs that are statistically significant in GWAS. This technique essentially shrinks all omitted SNPs to zero estimate and does not perform shrinkage on the effect size estimates of the included SNPs. PRS is often computed over a variety of thresholds given that the optimal \emph{p}-value threshold is a priori unknown and the target phenotype being evaluated for the given threshold while the forecast adjusted accordingly. This technique can be interpreted as a variable selection process which essentially executes the GWAS \emph{p}-value forward selection based on the size of the increment in the \emph{p}-value threshold.
\subsection{Bayesian approach in PRS analysis}
Bayesian approaches have been used to explicitly model pre-existing genetic architecture thereby accounting for the distribution of effect sizes with a prior that should improve the accuracy of a polygenic score. The main advantage of Bayesian-based PRS analysis is its ability to improve genomic risk prediction from summary statistics by taking into account linkage disequilibrium (LD) among markers \cite{So2017}.
\subsubsection{Empirical Bayes PRS (EB-PRS) method}
In general,  EB-PRS  method is a novel approach based on the Empirical Bayes theorem incorporating  information across markers to improve prediction accuracy \cite{Song2020}. EB-PRS method  aims at minimizing the prediction error by leveraging on the estimated distribution of effect sizes. Assuming the SNPs are independent, the optimal PRS value (in terms of achieving the best classification accuracy) is
\begin{eqnarray}
S=\beta^{T} X=\sum_{i=1}^{m}\beta_{i}X_{i},
\end{eqnarray}
where $m$ is the total number of genotyped SNPs. $X_i$ is the genotypic value and $\beta_i$ is the log-odds ratio (OR) of the $i$th SNP. The log-OR is a measure of the effect size defined in the following formula:
\begin{eqnarray}
\beta_{i}=\log\left(\frac{f_{i1}(1-f_{i0})}{f_{i0}(1-f_{i1})}\right),
\end{eqnarray}
where $f_{i0}$ and $f_{i1}$  are the reference allele frequencies among controls and cases respectively. If the SNP is not associated with disease, then  $\beta_{i}=0$.

In practice, the true values of effect sizes are usually unknown and need to be estimated from the data. Song \textit{et al.} \cite{Song2020}, in their method use the Empirical Bayes approach to estimate $\beta$ which is the minimizer of the Bayes risk under the distribution estimated from the data. The estimators can be derived directly from GWAS summary statistics. Compared to other improved genetic risk prediction methods \cite{Hu2017,Yang2015}, the EB-PRS method does not require external panels or datasets. While there are other methods which utilize effect size distributions for PRS value calculations, methods in this category such as \cite{So2017,Mak2017}  have no tuning parameters or external input. However, the EB-PRS method has a theoretical superiority compared with the existing methods in this category in terms of minimizing the prediction error. This method was applied to the following six complex disease traits;  asthma (AS), breast cancer (BC), celiac disease (CEL), Crohn's disease (CD), Parkinson's disease (PD) and type-2 diabetes (T2D) to illustrate the improved risk prediction performance in real data (Table 1). Furthermore, the authors recorded significant improvement when comparing the EB-PRS  method with all other methods that include the unadjusted PRS method, $P+T$, LDpred-inf, LDpred, Mak \textit{et al.}'s  \cite{Mak2017}. Although the EB-PRS method can achieve better performance without tuning any parameters and utilizing external information, its performance may be improved with external information e.g.  the LD information as used in LDpred. Also, in order to increase the prediction accuracy, Song \textit{et al.} \cite{Song2020} suggested that the EB-PRS method could be further improved by combining other available datasets in the future such as annotations or other GWAS summary statistics studying genetically correlated traits.

\subsubsection{Polygenic Risk Score-Continuous Shrinkage (PRS-CS) method}
The PRS-CS method is based on a Bayesian high-dimensional regression framework for polygenic modeling and prediction:
\begin{eqnarray}
Y_{N\times 1}=X_{N\times M}\beta_{M\times 1}+\epsilon_{N\times1},
\end{eqnarray}
where $N$ and $M$ denote the sample size and number of genetic markers respectively. $Y$ is a vector of traits while $X$ is the genotype matrix. $\beta$ is a vector of effect sizes for the genetic markers and $\epsilon$ is a vector of residuals. By assigning appropriate priors on the regression coefficients $\beta$ to impose regularization, additive PRS value can be calculated using posterior mean effect sizes.

Unlike LDpred \cite{Yang2015} and the normal-mixture model recently developed \cite{Zhang2018,LloydJones2019} which can incorporate genome-wide markers having varying genetic architectures with enhanced performance and flexibility, the PRS-CS method  utilizes a Bayesian regression framework and places a conceptually different class of priors—the continuous shrinkage (CS) priors—on SNP effect sizes \cite{Feng2019}. Continuous shrinkage priors allow for marker-specific adaptive shrinkage i.e. the amount of shrinkage applied to each genetic marker is adaptive to the strength of its association signal in GWAS which can accommodate diverse underlying genetic architectures.

Feng \& Smoller \cite{Feng2019} presented the PRS-CS-auto method, a fully Bayesian approach that enables automatic learning of a tuning parameter $\phi$, from GWAS summary statistics. Although analyses conducted from the Biobank indicate that for many disease phenotypes, the current GWAS sample sizes may not be large enough to accurately learn  $\phi$ and the prediction accuracy of PRS-CS-auto method  may be lower than PRS-CS method and LDpred. However, simulation studies and quantitative trait analyses suggest that PRS-CS-auto method can be useful when the training sample size is large or when an independent validation set is difficult to acquire. Although the PRS-CS method provides a substantial improvement over existing methods for polygenic prediction \cite{Yang2015}, current prediction accuracy of PRS value is still lower than what can be considered clinically useful. Much work is still needed to further improve the predictive performance and translational value of PRS methods.

Recent studies by \cite{MrquezLuna2018,Shi2016,Turley2019} argued that jointly modeling multiple genetically correlated traits and functional annotations in polygenic modeling are expected to increase the predictive performance of PRS methods.
\subsection{PRS methods based on Shrinkage of GWAS effect size estimates}
Since SNP effects are calculated with uncertainty and not all SNPs have an impact on the traits, unadjusted effect size estimates of all SNPs can lead to low-estimated PRS, with high standards error \cite{Choi2018}. Two shrinkage methods have been implemented to solve these problems:  shrinkage of the effect estimates of all SNPs by adapted statistical techniques, and use of \emph{p}-value filtering thresholds as the criterion for inclusion of SNPs.
\subsubsection{Shrinkage of the effect estimates of all SNPs by adapted statistical techniques}
PRS methods performing shrinkage of all SNPs \cite{Wray2014,Yang2015} typically apply shrinkage/regularisation techniques such as LASSO/ridge regression\cite{Wray2014}, or Bayesian approaches performing shrinkage by prior distribution specification\cite{Yang2015}.
Varying degrees of shrinkage may be accomplished under varying methods or parameter settings. The most suitable shrinkage to be implemented depends on the underlying mixture of distributions of null and true effect size which is likely to be a complex mixture of distributions that differ by traits. PRS estimation is usually tailored over a number of (tuning) parameters since the optimum shrinkage parameters are a priori unknown. For example, in the case of LDpred, it includes a setting for a fraction of causal variant \cite{Yang2015}.
\subsubsection{\emph{p}-value filtering thresholds as the criterion for inclusion of SNPs}
In this process, the PRS estimate includes SNPs with a GWAS  P-value below a certain level (e.g. \emph{p}-value $< 23^{-5}$) while all other SNPs are removed. This method shrinks all omitted SNPs to an estimated effect size of zero and does not perform shrinkage on the effect size estimates of the included SNPs. Since the optimum \emph{p}-value threshold is a priori unknown, PRS is computed over a range of thresholds, associated with each of the tested target trait and optimized appropriately for the prediction \ref{Fig1}). In systematic shrinkage techniques, this method is similar to tuning parameter optimization.
This technique is regarded as a parsimonious method of selection of variables. It is efficient in performing the forward selection of variables (SNPs) using GWAS \emph{p}-value with the sizes depending on the increment of P-value threshold. Therefore, only in the sense of this forward selection method is the chosen 'optimal threshold' defined; a PRS derived from another subset of the SNPs may be more predictive of the target trait. However, considering the fact that GWAS is focused on millions of SNPs, the number of subsets of SNPs that could be chosen for study is too high.

\subsection{Linkage Disequilibrium Control}
Usually, association studies in GWAS are done one-SNP-at-a-time \cite{Choi2018}. It combines with high genome-wide correlation structure, making it incredibly difficult to classify the independent genetic effects.
Though GWAS' power can be enhanced by leveraging on the results of several SNPs concurrently \cite{Loh2018}, provided that  raw data on all samples are available. Generally, researchers need to take advantage of standard GWAS (one SNP at a time) summary statistics for polygenic scoring. To estimate the PRS, there are two key options: (i) SNPs are clumped such that the retained SNPs are mostly independent of each other, (ii) all SNPs are included and the linkage disequilibrium (LD) between them is adjusted. 
In the 'normal approach' to polygenic scoring, option (i) is normally preferred, requiring \emph{p}-value thresholding, while option (ii) is commonly preferred in methods that incorporate conventional methods of shrinkage \cite{Mak2017,Yang2015} (see Table \ref{Tab2}). 
As for option (i) without clumping, some researchers tend to perform the \emph{p}-value thresholding method. Although breaking this presumption could lead to marginal losses in certain situations\cite{Mak2017}, Choi \textit{et al.} \cite{Choi2018} suggested that clumping be done where GWAS estimates of non-shrunk effect sizes are used. The standard method tends to work comparably to more advanced approaches \cite{Mak2017,Yang2015}. This could well be due to the clumping mechanism capturing conditionally independent effects. However, a critique of clumping is that for the elimination of SNPs in LD, researchers usually use an arbitrarily selected correlation threshold \cite{Wray2013}. Thus, no technique is without arbitrary features, this could be an area for the potential development of the classical method.

\clearpage
\begin{landscape}
\renewcommand{\arraystretch}{1.2}

\begin{longtable}{|p{2.5cm}|p{3cm}|p{2.5cm}|p{2cm}|p{7.5cm}|p{3cm}|} 
   \caption{Summary of polygenic risk score tools.}
    \label{table1}
 \\ \hline
    Tool &  Approach & Computational Platform & User Friendly & Functionality & Link \\
    \hline  
    \endfirsthead
    \multicolumn{6}{@{}l}{Summary of polygenic risk score tools\ldots continued.}\\ \hline
    Tool & Approach & Computational Platform & User Friendly & Functionality & Link\\
    \hline
    \endhead 
    LDpred\cite{Yang2015} & Bayesian Shrinkage Prior & Python & Difficult & Method that uses a prior on effect sizes and LD information from an external reference panel &{\tiny \url{ https://bitbucket.org/bjarni_vilhjalmsson/ldpred}} \\ [1em] \hline
    
    PRS-CS\cite{Feng2019} &  Bayesian regression framework & Python & Difficult & utilizes a high- dimensional Bayesian regression framework, by placing a continuous shrinkage (CS) prior on SNP effect sizes & \tiny \url{https://github.com/getian107/PRScs} \\ [5em] \hline
    
    EB-PRS\cite{Song2020} & Empirical Bayes approach & R & Difficult & A novel method that leverages information for effect sizes across all the markers & R CRAN \\ [1em] \hline
    
    AnnoPred\cite{Hu2017} & Bayesian Shrinkage Prior & Python & Difficult & A framework that leverages diverse types of genomic and epigenomic functional annotations &\tiny \url{https://github.com/yiminghu/AnnoPred}\\ [1em] \hline
    
    PRSice\cite{Euesden2015} & Clumping + thresholding (C+T) & R &  Difficult & for calculating, applying, evaluating and plotting the results of PRS analysis & \tiny \url{http://PRSice.info}\\ [1em] \hline
    
    PRSice2\cite{Choi2019} & Clumping +thresholding (C+T) & C++, R & Easy & an efficient and scalable software program for automating and simplifying PRS analyses on large-scale data & \tiny \url{http://PRSice.info}\\ [1em] \hline
    
    LDpred2\cite{Priv2020} & Bayesian Shrinkage & R & Difficult & A faster and more robust implementation of LDpred in R package bigsnpr & \tiny \url{https://privefl.github.io/bigsnpr/articles/LDpred2.html}\\ [1em] \hline
    
    BSLMM\cite{Yang2020} & Bayesian sparse linear mixed model & R & Difficult & Prior specification for the hyper-parameters and a novel Markov chain Monte Carlo algorithm for posterior inference & \tiny \url{http://stephenslab.chicago.edu/software.html} \\[1em] \hline
    
    BayesR\cite{LloydJones2019} & Hierarchical Bayesian Mixture Model & Fortran & Difficult & Bayesian mixture model that simultaneously allows variant discovery, estimation of genetic variance explained by all variants. & \tiny \url{https://github.com/syntheke/bayesR}\\ [5em] \hline
    
    DPR software\cite{Zeng2017} & Latent Dirichlet process regression model & C++ & Easy & Dirichlet process regression to flexibly and adaptively model the effect size distribution & \tiny \url{http://www.xzlab.org/software.html}\\ [1em] \hline
    
    SMTpred\cite{Maier2018}& & Python & Difficult & Combines SNP effects or individual scores from multiple traits according to their sample size, SNP-heritability ($h^2$) and genetic correlation ($r_G$). & \tiny \url{https://github.com/uqrmaie1/smtpred} \\ [1em] \hline
    
    Lassosum \cite{Mak2017}& Penalised Regression & R & Difficult & A method for constructing PGS using summary statistics and a reference panel in a penalized regression framework & \tiny \url{https://github.com/tshmak/lassosum}\\ [1em] \hline
    
    Plink\cite{Chang2015} & \emph{p}-value thresholding approach & C/C++ & Easy  & Open-source C/C++ toolset for genome-wide association studies (GWAS) and research in population genetics & \tiny{\url{http://zzz.bwh.harvard.edu/plink/}}\\
\hline

\end{longtable}
\end{landscape}

\begin{table}[h]
\begin{adjustwidth}{-2.25in}{0in}
    \caption{Comparison of different approaches for performing PRS analyses.}

    \centering
    \begin{tabular}{|p{2.1cm}|p{3cm}|p{3cm}|p{3cm}|p{3cm}|}
    \hline
    \multirow{2}{*}{Key Factors} & \multicolumn{4}{c|}{Approaches}\\ [3ex] \cline{2-5}
        & \emph{p}-value thresholding  with clumping & Penalised Regression & Clumping + Thresholding (C+T) & Bayesian Shrinkage Prior\\  [3ex] \hline  
        Controlling for Linkage Disequilibrium & 	N/A	          & LD matrix is integral to algorithm & Clumping	& Shrink effect sizes with respect to LD \\ \hline
        Shrinkage of GWAS effect size estimates & 	\emph{P}-value threshold & LASSO, Elastic Net, penalty parameters Bayesian	&  P-value threshold Standard &	Prior distribution, e.g. fraction of causal SNPs \\  [3ex] \hline
    \end{tabular}
    \label{Tab2}
    \end{adjustwidth}
\end{table}

\section{PRS and Population structure/Global heterogeneity}
The key cause of confounding in GWAS (post-QC) is population structure, hence the possibility of false-positive results \cite{Choi2018}. In general, structure in mating patterns induces structure in genetic variation,  closely associated with geographic location. Furthermore, environmental risk factors may be organized in a similar manner, creating the possibility for correlations between certain genetic variations and the characteristics examined that are confounded by, for example, location \cite{Price2006,Astle2009}. Usually, this issue is solved in GWAS by modifying the principal components (PCs)\cite{Price2006} or by using mixed models \cite{Price2010}. However, population composition in the PRS study presents a possible greater issue since a significant number of null variants usually are included in PRS estimation. For example, allele frequencies are systematically different between the base and target data that can be obtained from genetic drift or genotyped variant \cite{Kim2018}. In addition, there is a danger that variations in null SNPs may result in the correlation between the PRS and target traits if the distributions of the environmental risk factors for the phenotype vary in both (base and target data) – both highly probable in most PRS studies. Even if the GWAS had completely regulated its population structure, confounding is possibly reintroduced. Correlated variations between the base and target data in allele frequencies and risk factors are not taken into consideration.
The regulation of structure in the PRS study should be adequate to prevent false-positive if the base and target samples are drawn from the same or genetically similar populations. Choi \textit{et al.} advised that care should be taken, provided that there are drastic variations between populations in the distribution of PRS \cite{Kim2018,Martin2017,Duncan2018}. Such observations do not indicate large differences between populations in aetiology while genuine differences are likely to contribute due to geographical, cultural and selection pressure variations. It challenges the accurate use of base and target data from different populations in PRS studies that do not tackle the problem of possible uncertainty generated by geographical stratification\cite{Martin2017}. It is therefore important to be mindful that, by exploiting large sampling sizes, extremely significant effects can be obtained due to subtle confounding. Population structure issues are as significant as the variations between individuals in the base and target populations in genetics and the environment. In the coming years, the topic of generalizability of PRS methods across populations is expected to be an active field \cite{Duncan2018,MrquezLuna2017}.

\section{PRS tools}
The next sections would provide an example of some PRS tools that are commonly used to perform PRS analysis. 
\subsubsection{Linkage Disequilibrium Pred (LDpred)}{\label{LDPred}}
This method infers the posterior mean effect size of each marker by using a prior on effect sizes and LD information from an external reference panel \cite{Yang2015}. LDpred calculates the posterior mean effects from GWAS summary statistics by conditioning on a genetic architecture prior and on LD information from a reference panel. The inner product of these  is re-weighted and the test-sample genotypes is the posterior mean phenotype, and under the model assumptions and available data, posterior mean phenotype is an optimal (minimum variance and unbiased) predictor. The prior of the effect sizes is a point-normal mixture distribution which allows for non-infinitesimal genetic architectures. The prior has two parameters; the heritability, explained by the genotypes and the fraction of causal markers i.e. the fraction of markers with non-zero effects. The heritability parameter is estimated from GWAS summary statistics and accounts for sampling noise and LD \cite{Finucane2015}.

By applying LDpred to five diseases: Sczherhernia (SCZ), Muscular dystrophy (MS), BC, Type II diabetes (T2D) and Coronary artery disease  (CAD) for which the GWAS summary statistics for large sample sizes ranging from 27,000 to 86,000 individuals and raw genotypes for an independent validation dataset, LDpred outperforms the approach of pruning followed by thresholding \cite{So2017}, particularly at large sample sizes. For instance, in a large dataset of schizophrenia and multiple sclerosis, the predicted $R^2$ increased from 20.1\% to 25.3\% and from 9.8\% to 12.0\%, respectively.  In another test, LDpred was applied to predict SCZ risk in non-European validation samples of both African and Asian descents. Although prediction accuracies were lower in absolute terms,  similar relative improvements were observed for LDpred over other methods.

LDpred is a popular and powerful method for deriving polygenic scores based on summary statistics and a Linkage Disequilibrium (LD) matrix only \cite{Vilhjlmsson2015}. It assumes there is a proportion p of variants that are causal. However, LDpred has several limitations that may result in limited predictive performance. The non-infinitesimal version of LDpred, a Gibbs sampler, is particularly sensitive to model misspecification when applied to summary statistics with large sample sizes. It is also unstable in long range LD regions such as the human leukocyte antigen (HLA) region of chromosome 6. This issue has led to the removal of such regions from analyses \cite{LloydJones2019,MrquezLuna2018} which is unfortunate since this region of the genome contains many known disease-associated variants, particularly with autoimmune diseases and psychiatric disorders \cite{Mokhtari2016, Matzaraki2017}. In a recent development, a new version of LDpred that addresses these issues while markedly improving its computational efficiency was presented  by Privé \textit{et al.} \cite{Priv2020}. This is a faster and more robust implementation of LDpred in the R package bigsnpr.
\subsection{LDpred2}
A new version of LDpred, LDpred2, has  a \textit{sparse} option that can learn effects that are exactly $0$ and an \textit{auto} option that directly learns parameters from data. LDpred is widely used and has the potential to provide polygenic models with good predictive performance \cite{Khera2018}. Yet, it has some instability issues that have been pointed out by Marquez-Luna \textit{et al.} \cite{MrquezLuna2018}  and by Lloyd-Jones \textit{et al.} \cite{LloydJones2019} and likely contributed to the discrepancies in reported prediction accuracies \cite{Choi2019,Ge2019}. For instance, LDpred1 performs poorly in the simulations where causal variants are in the HLA region. In contrast, LDpred2 performs very well. It uses a window size of 3 centiMorgan (cM), which is larger than the default value used in LDpred1 and enables LDpred2 to work well even when causal variants are in long-range LD regions. In another scenario, LDpred2-auto which automatically computes values for hyper-parameters $p$ and $h^2$, equally performs well compared to other LDpred2 models in simulations but does not perform well for some of the real traits. Typically, Type 1 diabetes (T1D) is mainly composed of large effects in the HLA region because summary statistics have a small sample size. It is unknown why LDpred2-auto performs poorly specifically for pure red cell aplasi (PRCA). More studies need to be performed to understand the poor results of LDpred2-auto in these two cases.

\subsection{PRSice}
In 2015, Euesden \textit{et al.} \cite{Euesden2015} developed the first dedicated PRS analysis software, naming the resulting methodology PRSice. PRSice is written in R, with wrappers for bash data management scripts and PLINK-1.9 to minimize computational time (Table 1). Considering $n$ individuals from the 'target phenotype' data set using a list of $m$ SNPs, the genotypes have some effect (or not) on the 'base phenotype'. The base and target phenotype may be the same if assessing the shared genetic overlap of a phenotype between samples/populations. These genotype effects can be estimated from a univariate regression on the base phenotype for each SNP, such as from a genome-wide association study (GWAS). In such a GWAS for a SNP $i$, where $i$ = 1, 2, ..., $m$, a \emph{p}-value, $P_i$, is calculated for the association between the SNP genotypes, $G_i$,$j = \{0,1,2\}$ for individual $j$ where $j$ = 1, 2, … , $n$ and the phenotype. Under the usual additive assumption made in GWAS, a corresponding effect size is estimated by $\beta_i$ for the effect of a unit increase in genotype $G_{ij}$, on the phenotype. SNPs are generally selected for inclusion in a PRS value based on the degree of evidence according to \emph{p}-value for their association with the base phenotype in a GWAS – SNP $i$ will be included in a PRS calculation if $P_i$ is smaller than a threshold, $P_T$. PRS values  are typically calculated at a number of different \emph{p}-value thresholds, $P_T$.

At threshold $P_T$, the PRS value for individual $j$ can be calculated as:
\begin{eqnarray}
PRS_{PT,j}=\sum_{i=1}^{m}\beta_{i}G_{i,j}.
\end{eqnarray}
The PRS value is calculated across all individuals giving $n$ scores per threshold, $P_T$. The
association between these PRS values and the target phenotype can then be evaluated in an appropriate regression model (depending on the data type of the target phenotype, e.g. linear regression if the phenotype is continuous). PRSice tool has been developed to fully automate PRS analyses, substantially expanding the capability of PLINK-1.9 \cite{Chang2014}. In real data, there is usually some missing genotype data unless genotypes have already been imputed. PLINK-1.9 imputes any missing data according to mean allele frequencies. However, it is not equipped to handle very large data sets, and a more memory-efficient approach is used in its advanced version, PRSice-2.
\subsection{PRSice-2}
PRSice-2 which is an enhancement of PRSice, handles both genotyped and imputed data, provides empirical association \emph{p}-values free from inflation due to overfitting, supports different inheritance models and evaluates multiple continuous and binary target traits simultaneously \cite{Choi2019}. This method streamlines the entire PRS analysis pipeline without generating intermediate files and performs all the main computations in C++, leading to a drastic speed-up in run time and reduction in memory burden. Furthermore, using best-guess genotypes (BGEN) imputation format, PRSice-2 can directly process the BGEN imputed format and convert to either best-guess genotypes or dosages when calculating the PRS value without generating a large intermediate file. While PRS values based on best-guess genotypes are calculated as for genotyped input, dosage-based PRS values are calculated as

\begin{eqnarray}
PRS=\sum_{i}^{m}\beta_{i}\left(\sum_{j}^{2}w_{ij}X_{j}\right).
\end{eqnarray}

Where $\omega_{ij}$ is the probability of observing genotype $j$,where $j \in \{0,1,2\}$ , for the $i^{th}$ SNP; $m$ is the number of SNPs; and $\beta_i$ is the effect size of the $i^{th}$ SNP estimated from the relevant base genome-wide association study (GWAS) data. A simulation study has been used to compare the performance of PRSice-2 to alternative polygenic score software lassosum\cite{Mak2017} and LDpred \cite{Yang2015} in terms of run time, memory usage and predictive power on servers equipped with 286 Intel 8168 24 core processors at 2.7 GHz and 192 GB of RAM.

Based on the simulation results, PRSice-2 showed best performance in all settings, significantly faster than lassosum and LDpred. Specifically, PRSice-2 can complete the full PRS analysis on 100,000 samples within 4 minutes,  which is 179 times faster than the 10 hours required by lassosum and 241 times faster than the 13 hours 27 minutes required by LDpred. Similarly, PRSice-2 requires significantly less memory than lassosum and LDpred, requiring $<$500 MB of memory for 100,000 samples as opposed to 11.2 GB required by lassosum and 45.2 GB required by LDpred.

Another case study  compared the predictive power of PRSice-2 to lassosum and LDpred for quantitative traits with heritability of 0.2, base sample size of 50,000 and target sample size of 10,000. PRSice-2 has comparable predictive power to lassosum and LDpred, typically generating PRS values with predictive power higher than those of LDpred but not as high as lassosum. The details of the simulation code can be found here (\href{https://github.com/choishingwan/PRSice-paper-script}{simulation code}), for others to inspect and repeat the analyses. While PRS values generated by PRSice-2 do not seem to fully optimize predictive accuracy, the simple approach and typically fewer SNPs exploited allow for easier interpretation of the results compared with methods that use all SNPs \cite{Cecile2019}.

\subsection{Lassosum}
It is an alternative method that uses summary statistical data to estimate PRS, and takes LD into account by using reference panels \cite{Mak2017} on the basis of the commonly used LASSO and elastic net regression \cite{Tibshirani1996,Zou2005}. Consider the linear regression given below:
\begin{eqnarray}
y=X\beta + \epsilon.
\end{eqnarray}
For which $X$ represents a data matrix of \emph{n}-by-\emph{p}, and $y$ denotes a vector of the observed outcome. LASSO is a commonly used method for deriving $\beta$ estimates and y predictors, especially in cases where p is high and where it is rational to conclude that many $\beta$ are 0. By minimizing the objective function, LASSO also obtains estimates of $\beta$ given $y$ and X.
To test the efficiency of lassosum relative to LDpred, simulation studies\cite{Vilhjlmsson2015} were carried out using summary statistics for which LD was accounted for and Welcome Trust Case Control Consortium (WTCCC) Phase 1 data for seven diseases. The outcome of LDpred, lassosum, and simple soft-thresholding (setting s = 1 in lassosum) was comparable for most of the diseases in the WTCCC dataset, except for T1D, where lassosum seem to outperform LDPred. The performance of LDpred and lassosum was comparable when the number of causal SNPs was 1,000 and the sample size was 11,200 for the simulated phenotypes, and both were superior to soft thresholding. However, LDpred's performance was considerably reduced when the sample size was halved. The lassosum was not influenced in the same way when reducing the sample size by halve. All methods performed equally when the number of causal SNPs was 25,000 and the sample size was 11,200.
The fact that summary statistics can be confounded by population stratification and population heterogeneity, makes real-life application of PRS difficult. However, these problems in the lassosum design were not considered. One possible issue with the use of meta-analytical summary statistics is that the original data produced by the summary statistics is an amalgamation of datasets around the world with correction for population stratification. Possibly, there is no one homogenous dataset suitable as a reference panel. Further research is therefore required to explain what is the best approach here. 

Schork \textit{et al.} \cite{Schork2013} have demonstrated that different genome regions have different false discovery rates, thus different chances of being causally correlated with a phenotype. Therefore, genome annotation information can theoretically be used to enhance the performance. Similarly, it is possible to utilize the fact that certain phenotypes have common genetic determinants (pleiotropy) to improve PRS.

\subsection{PLINK SOFTWARE (Second-generation PLINK)}
PLINK 1 is an open-source C/C++ tool set for performing genome-wide association studies (GWAS) and research in population genetics. However, the steady growth of data from imputation and whole-genome sequencing studies called for an urgent need for faster and scalable implementations of its key functions. In addition, GWAS and population-genetic data now frequently contain genotype likelihoods, phase information, and/or multiallelic variants, none of which can be represented by PLINK 1's primary data format. 

To address these problems, Chang \textit{et al.}  \cite{Chang2015} developed a second-generation codebase for PLINK. The first major release from this codebase, PLINK 1.9, introduces extensive use of bit-level parallelism, $O(\sqrt{n})$ -time/constant-space Hardy-Weinberg equilibrium calculation and Fisher's exact tests and many other algorithmic improvements. In combination, these changes accelerate most operations by 1-4 orders of magnitude and allow the program to handle data sets too large to fit into RAM. 

PLINK 1.9's core functional domains are unchanged from that of its predecessor, and it is usable as a drop-in replacement in most cases with no changes to existing scripts. To support easier interoperation with newer software, features such as the import/export of VCF and Oxford-format files and an efficient cross-platform genomic relationship matrix (GRM) calculator have been introduced. Most pipelines currently employing PLINK 1.07 can expect to benefit from upgrading to PLINK 1.9.  Despite its computational advances, PLINK 1.9 can still be an unsatisfactory tool for working with imputed genomic data due to the limitations of the PLINK 1 binary file format. To address this issue, the authors designed a new core file format in PLINK 2.0 capable of representing most of the information emitted by modern imputation tools.
\subsection{PRS tools that are applicable to diverse populations}
Applying PRS analysis  for multi-ethnic groups is still limited. However, novel  PRS methods have been developed to address the applicability of PRS analysis across ethnic groups.
\subsubsection{Multi-ethnic PRS analysis}
Multi-ethnic PRS  analysis is a new PRS  approach that combines PRS analysis based on two distinct populations \cite{MrquezLuna2017}. For instance, multi-ethnic PRS analysis could merge PRS analysis based on European training data with PRS  analysis based on training data from another population.  The multi-ethnic PRS approach computes PRS value  given a target individual with genotypes $g$  as follows:  
\begin{eqnarray}
PRS=\sum_{i=1}^{M}{ \hat{b}_{i} g_{i}},
\end{eqnarray}

where $M$ is the total number of individual's genetic markers, and $\hat{b}_{i}$  is an estimate of  effect sizes. For a multi-ethnic PRS analysis, this approach uses a  linear combination of the two distinct PRS values and applying mixing weights parameters $\alpha_{i}$. 
\subsubsection{Linear Unbiased Predictors (BLUP)}
PRS analysis could be molded using the well-known approach of best linear unbiased predictors (BLUP) \cite{Chen2015}.  BLUP is used to consider and linearly model both random effects and fixed effects. It is also known as genomic best linear unbiased prediction (gBLUP). \cite{Clark2013}. BLUP/gBLUP estimates PRS values using the following formula 
\begin{eqnarray}
PRS=X\beta + g + \epsilon,
\end{eqnarray}
where $\beta$ is a vector of fixed effects,  $g$ is the total genetic effects of the training samples and $\epsilon$ are the normally distributed   residuals. To evaluate the fixed effects,  BLUP considers  an individual GWAS  indicator, the top 5 principal components  (PCs) derived with all samples together and/or a list of the significant SNPs. The BLUP approach is a computationally  efficient algorithm. However, the limitation of BLUP arose due to its requirement of the Individual-level genotype data.  BLUP has been implemented in GCTA software (Genome-wide Complex Trait Analysis) \url{https://cnsgenomics.com/software/gcta/#Overview}. Moreover, it has been extended to XP-BLUP to model PRS values for admixed populations\cite{Clark2013}. Also, BLUP has been extended to MultiBLUP to include multiple random effects \cite{Speed2014}.
\subsubsection{Genetic Risk Scores Inference (GeRSI)}
GeRSI uses mixed models by combing fixed-effects models and random-effects models for controlling population structure \cite{Golan2014}. GeRSI performs Gibbs sampling to estimate individuals' genetic risk score given the case-control study's genotypes under a random-effects model. GeRSI proposed conditional distributions of the genetic and environmental using the standard liability-threshold model. However, the limitation of GeRSI is that it requires individual-level genotypes which will not be available to many bioinformaticians. 
\subsubsection{Cross-population BLUP (XP-BLUP)}
XP-BLUP is an extension of the BLUP  method that could be applied to trans-ethnic populations \cite{Clark2013}.  XP-BLUP utilizes trans-ethnic information to improve PRS value  predictive accuracy in minority populations. XP-BLUP combines the linear mixed-effects model (LMM) of the GeRSI method with the BLUP  method. 

\section{The predictive power of PRS analysis}
Within the current literature that addresses the statistical power of the PRS analysis, most of these articles consider the sample size as a milestone to power the PRS analysis. For instance, in 2013, Dudbridge estimated the predictive power of the polygenic score using results from several published studies \cite{Dudbridge2013}.  Dudbridge has concluded that all published studies with a significant association of PRS values are statistically well-powered. Also,  Dudbridge pointed out that the accuracy of the PRS analysis depends only on the size of the initial sample (training sample). Furthermore, he provided a mathematical model to estimate the statistical power of PRS value as a function of sample size.  In 2014, Middeldorp \textit{et al.} \cite{Wray2014} suggested performing PRS analysis on a sample size of 2000 individuals is good enough to obtain a statistically powered PRS value.  However, Dima and  Breen in 2015 \cite{Dima2015} demonstrated that a sample size of 1500 is enough to increase the predictive power to a  statistically significant point. However, they stated that the predictive power of polygenic risk scores is not good enough for  clinical applications but it could be used as a biomarker for traits of interest within individuals. Recently, in 2017,  Krapohl \textit{et al.} \cite{Krapohl2017} introduced a multi-polygenic score that is capable of increasing the predictive power of PRS analysis.

Regarding the relative accuracy of PRS values across ancestries, Yengo \textit{et al.} \cite{Wang2020} proposed a theoretical model to estimate the relative accuracy of PRS value across ancestries.  Their method utilizes the frequencies of the minor alleles (MAF) in the two populations,  the LD between the causal SNPs and the heritabilities.  The authors assumed that causal variants are shared across ancestries. However,   their effect sizes might vary.  Based on their model,  \cite{Wang2020}  concluded that LD and MAF differences across ancestries explained 70-80\% of the loss of relative accuracy of European-based PRS value in African ancestry.

\section{PRS analysis on the African population}

The approach of the PRS analysis is still not applied to study traits in the African population. For instance,  upon searching  PubMed using the keywords: "polygenic risk" + "African", only 78 hits was obtained.  This number represents about 5.45\% of total hits  that were obtained without using the keyword "African".  The traits studied using PRS analysis in the African population include  types 1 \& 2 diabetes mellitus,  depression,  ischemic stroke, schizophrenia, sarcoidosis , alzheimer's disease, obesity, insomnia disorder, post-traumatic stress and cancer. The following paragraphs will demonstrate the selected  PRS studies done on the African sub-saharan populations.  Also, we will highlight their outcomes.

In 2020,  Ekoru and his colleagues  investigated the genetic risk scores for cardiometabolic traits in several African ancestries, including  sub-saharan African populations \cite{Ekoru2020}. They concluded that the predictive power of the risk score is limited in the African ancestry populations. They stated that this limitation is due to the insufficient diversity among their samples of genomic discovery. Therefore, they adjusted for ancestry-derived principal components to obtain  up to 5-fold and 20-fold higher predictive power. However, they observed that the predictive power of genetic risk scores was higher in the African Americans (n=9139) and the  European Americans (n=9594) relative to the sub-saharan African populations (n=5200). Based on their outcome, Ekoru and his colleagues concluded that PRS  analysis performs poorly in   sub-saharan African populations. Also, they recommended paying attention to the representation of multi-ethnic populations in genomic studies to improve the power of the genetic risk scores. 

In 2020, Hayat and her colleagues investigated the genetic associations between serum low LDL-cholesterol levels and selected genetics variants in sub-saharan African of four countries; Kenya, South Africa, Ghana and Burkina Faso \cite{Hayat2020}. Using 1000 genomes data from the African populations, they  selected four genes for their investigation (\textit{LDLR}, \textit{APOB}, \textit{PCSK9}, and \textit{LDLRAP1}). They performed genotyping of 19 SNPs using 1000 participants in the Human Heredity and Health in Africa (H3Africa) AWI-Gen Collaborative Center (Africa, Wits-INDEPTH Partnership for GENomic studies). Although they used a limited number of variants, their outcome showed significant associations of these SNPs with lower LDL-cholesterol levels in sub-saharan Africans. 

In 2020, Cavazos and Witte have proposed the inclusion of variants discovered from various populations to improve PRS transferability for diverse populations \cite{Cavazos2020}. They used both simulated data for the Yoruba group for the sub-saharan African population and European populations. They tested their findings on real data consisting of diabetes-free training samples of European ancestry (\emph{n} = 123,665) and African descent (\emph{n} = 7,564). They evaluated  performance  of PRS analysis using genotype and phenotype data for a test (predictive) sample of European ancestry (\emph{n} = 394,472) individuals of African origin from the UK Biobank (\emph{n} = 5,886). Based on their findings, they concluded that incorporating variants selected from the European population will limit the accuracy of PRS values in non-Europeans populations including African communities. Also, they commented on the need for diverse GWAS data to improve PRS accuracy across populations.

In 2017, Marquez-Luna \textit{et al.} \cite{MrquezLuna2017} proposed a multi-ethnic PRS analysis to improve risk prediction in diverse populations including the African community. To overcome the lack of enough training data for the African populations, the authors combined the training data to involve data from European samples and training data from the target population. As the authors did not state whether they used sub-saharan African communities, we  did not include their study.   However,  this	highlights the challenge of performing PRS analysis in sub-saharan African	populations due to lack of enough  training data.

In 2017, Vassos \textit{et al.} had examined  PRS values in a group of individuals with first-episode psychosis \cite{Vassos2017}. For the control sample, they combined African European  (\emph{n} = 70) and a sample of sub-saharan African ancestry (\emph{n}=828). Their finding showed that PRS value was more potent in  Europeans i.e. 9.4\% discriminative ability than  in Africans i.e. only 1.1\% discriminative ability in Africans.

Moreover, PRS analysis has been applied to investigate the risk score for prostate cancer.   Prostate cancer is considered a complex genetic disease with high heritability and disproportionally affects men of African descent\cite{Rebbeck2017}. In a study to predict the risks of prostate cancer in urban African populations, involving seven African study sites as well as European men from the 1000 Genomes Project. It was determined that risks of prostate cancer are much more significant for African genomes than European genomes (\emph{p}-value $<$ 2.2 x 10-16, Wilcoxon rank-sum test). This continental level pattern is consistent with public health data\cite{Bray2018}. A further investigation done by the team of MADCaP (Men of African Descent and Carcinoma of the Prostate Consortium)  to study sites portrayed a substantial amount of overlap in the PRS distributions of different African populations. Based on their findings, the investigators of MADCaP observed within-continent heterogeneity for the predicted risk of prostate cancer. Their findings showed that individuals from Dakar, Senegal have lower predicted risks of prostate cancer than other African study sites while individuals from Abuja, Nigeria have higher predicted risks of prostate cancer than other African study sites. The MADCaP team concluded  that allele frequency differences at common disease-associated loci could contribute to population-level differences in prostate cancer risk.
\section{Challenges of PRS analysis for the African population}
Many PRS methods have been developed and applied to test the risk score of individuals. Nevertheless, PRS analysis has not yet been used in the clinical field for the African population. There are still many limitations and challenges regarding the application of PRS analysis in the African population.  One of these challenges is lack of sufficient data to perform PRS analysis. For instance, querying the a term "sub-saharan" in the GWAS Catalog repository, the search resulted in only 70 publications out of 4,628 papers. Considering that several publications might use the same GWAS data, we could affirm that more GWAS experiments need to be done for the sub-saharan African population.   This might be due to lack of infrastructure and  funding to perform GWAS experiments in many countries in Africa. Also,  such restrictions might be due to the fact that  many African scientists  are still focusing their research on  infectious diseases like   malaria, tuberculosis and HIV. However, providing funding priority for infectious diseases is necessary for the African communities as they account for a higher mortality rate in Africa.

Due to lack of enough training and test data sets, some scientists choose to use training data from European samples which results in decreasing PRS prediction accuracy. Therefore, PRS analysis is not widely applied to clinical research in Africa.  Moreover, considering the diversity among African population, the model used for PRS analysis might not work for African sub-populations.  Therefore, our future direction would be to develop African-specific PRS methods that combine different sources of information.  

Another challenge in performing and applying PRS  analysis in the African population is the lack of long-term funds for GWAS experiments.  Therefore, African state authorities should be made aware of this challenge, so as to make more funds available for genomic research.   Howbeit, the funds should not be limited to the research institutes and principal investigators alone, they should equally be directed towards the provision of scholarships (postgraduate programs like PhD) and financial aids for young African researchers.  We have some promising African research consortiums (e.g. the pan African Bioinformatics Network for the Human Heredity and Health in Africa (H3ABioNet, {\it h3abionet.org}) and the Human Heredity and Health in Africa (H3Africa, {\it h3africa.org})) that are contributing in this regard. However, their funds come from outside Africa. There are also regional African efforts like the World Bank funded Africa Center of Excellence (ACE) I and the one following this, the ACE Impact, but these initiatives consist of few genomic research projects. A follow up project to the H3Africa, dedicated to data science health research, entitled Harnessing Data Science for Health Discovery and Innovation in Africa (DS-I Africa) will soon commence.

Moreover, lack of a pan-African genomic advisory board remains another challenge for genomic research in Africa including PRS analysis. The  existence of  such a research advisory board would help research transparency and  establish  ethical guidelines to perform genomic research. This could open the window to get more grants from funding agencies such as the National Center for Biotechnology Information (NCBI).  It is clear that without a rigorous ethical guide and transparency policies, it is hard to get long-term funds. 

\section{PRS analysis on type-2 diabetes, breast and prostate cancers}
PRS analysis has been successfully applied to estimate and identify individuals with genetic risk for many biological traits especially type-2 diabetes, breast cancer and prostate cancer (See supplementary file S1). Most of these studies provide significant evidence of the success of PRS analysis in identifying patients who are at high risk of developing disease complications. Hence,  the primary strength of  PRS analysis is its capability of stratifying individuals based on their probability of developing a disease. Also, the biological power of PRS analysis arose due to its potential capacity to identify therapeutic and genomic pathways for type-2 diabetes, breast cancer and prostate cancer.  Moreover,  applying PRS  analysis on these traits showed that PRS results are reproducible in the European population.

Nonetheless, one weakness of applying PRS analysis on these traits is its limited ability in detecting the false-positive results.  Also, it is observed that most PRS studies are only available for European ancestries. Therefore, we can not apply them to non-European communities. In addition, performing PRS analysis on a sizeable multi-ethnic data is indispensable for obtaining more accurate PRS values across populations. Furthermore, the possibility of applying PRS outcomes for personalized medicine requires robust validation procedures before broad clinical applications for multi-ethnic communities. 

\section{Conclusion and future research}
There are several approaches under the umbrella of PRS analysis. GWAS are conducted on finite samples extracted from particular subsets of the human population. Moreover, the SNP effect size estimates are some combination of true effect and stochastic variation, thus producing 'winner's curse' among the top-ranking associations, and the estimated effects may not generalized well to different populations. Furthermore, the correlation complicates the aggregation of SNP effects across the genome, therefore to apply PRS analysis across ethnic groups, 'Linkage Disequilibrium' (LD) holds the key. Thus, critical factors in the development of methods for calculating PRS values are
\begin{itemize}
    \item The potential adjustment of GWAS estimated effect sizes  e.g. via shrinkage and incorporation of their uncertainty;
\item The tailoring of PRS values to target populations; and
\item The task of dealing with LD.
\end{itemize}
As members of the H3Africa consortium and the associated bioinformatics consortium, H3ABioNet (see \url{h3abionet.org} and \url{https://sysbiolpgwas.waslitbre.org}), we are working to extend existing methods to be applicable to African populations. Also,  one  future direction will be to develop  an African-specific PRS method that combines the different sources of information.  
\section{Acknowledgements}
Research reported in this publication is supported by the National Human Genome Research Institute (NHGRI), Office Of The Director, National Institutes Of Health (OD) under award number U24HG006941 and the World Bank funding for the ACE Impact projects. The content is solely the responsibility of the authors and does not necessarily represent the official views of the National Institutes of Health and the World Bank. Our special thanks to Kalyani Dhusia for her editorial assistance.
\section{Competing interests}
The authors declare that they have no competing interests
\section{Organization Description}
H3ABioNet is a pan-African bioinformatics network comprising 28 bioinformatics research groups distributed amongst 16 African countries and 2 partner institutions in the USA. The consortium supports H3Africa researchers and their projects whilst developing bioinformatics capacity within Africa

\nolinenumbers

\bibliographystyle{unsrt}
\bibliography{Adametal-Polygenic Risk Score-in-Africa-Population-Final.bib}

\end{document}